\documentclass{article}
\usepackage{spconf,amsmath,amsfonts,epsfig}
\usepackage{multirow}

\title{
An Ensemble Approach for Compressive Sensing with Quantum Annealers
}

\name{
	Ramin Ayanzadeh, Milton Halem and Tim Finin
	\thanks{This research has been supported by NASA grant (\#NNH16ZDA001N-AIST 16-0091), NIH-NIGMS Initiative for Maximizing Student Development Grant (2 R25-GM55036), and the Google Lime scholarship. We would like to thank the D-Wave Systems management team, namely Rene Copeland, for granting access to the D-Wave 2000Q quantum annealer.}
}

\address{Department of Computer Science and Electrical Engineering\\
	University of Maryland, Baltimore County\\
	1000 Hilltop Cr., Baltimore, MD 21250\\
\texttt{\{ayanzadeh, halem, finin\}@umbc.edu}}
\begin{document}
%
\maketitle
\begin{abstract}
We leverage the idea of a statistical ensemble to improve the quality of quantum annealing based binary compressive sensing. 
Since executing quantum machine instructions on a quantum annealer can result in an excited state, rather than the ground state of the given Hamiltonian, we use different penalty parameters to generate multiple distinct quadratic unconstrained binary optimization (QUBO) functions whose ground state(s) represent a potential solution of the original problem. 
We then employ the attained samples from minimizing all corresponding (different) QUBOs to estimate the solution of the problem of binary compressive sensing. 
Our experiments, on a D-Wave 2000Q quantum processor, demonstrated that the proposed ensemble scheme is notably less sensitive to the calibration of the penalty parameter that controls the trade-off between the feasibility and sparsity of recoveries.
\end{abstract}
\section{Introduction} \label{sec:intro}
Compressive sensing (a.k.a. compressed sensing, compressive sampling or sparse sampling) is a recent sensing approach that exploits the sparsity of signals through optimization methods and reconstructs sparse (and compressible) signals from far fewer samples than the imposed rate by the sampling theorem \cite{candes2004robust,donoho2006compressed,candes2004near,mousavi2019survey}.
From an application point of view, compressive sensing has demonstrated outstanding performance where: (a) we are restricted by the factor of energy consumption on sensing side (e.g., wireless sensor networks); (b) we are limited to use few sensors (like hyper-spectral wavelengths); (c) sensing is time-consuming (namely medical imaging); or (d) measurement/sensing is too expensive (e.g., high-speed analog-to-digital-convertors) \cite{rani2018systematic} \cite{ayanzadeh2019compressive}.

The original problem of compressive sensing (a.k.a. the $\ell_0$-norm sparse recovery) aims to recover a sparse signal $\mathbf{x} \in \mathbb{R}^N$ from a given measurement vector $\mathbf{y} \in \mathbb{R}^m$ such that $\mathbf{y}=A\mathbf{x}$ and the measurement matrix $A \in \mathbb{R}^{m \times N}$ with $m \ll N.$ 
This problem can have infinite solutions, and compressive sensing guarantees the uniqueness of sparse solutions under various conditions \cite{baraniuk2008simple,foucart2013mathematical,mousavi2019survey}. 

Let $\| \mathbf{x} \|_0$ denotes the sparsity level of $\mathbf{x}$ (i.e., the number of nonzero entries of $\mathbf{x}$). 
We can represent the ultimate goal of compressive sensing  as
\begin{equation}
	\label{eqn:cs_ell_0}
	\min_{\mathbf{x} \in \mathbb{R}^N}{\|\mathbf{x}\|_0} \quad \textnormal{s.t.} \quad \mathbf{y}=A\mathbf{x}.
\end{equation}
This problem is NP-hard \cite{muthukrishnan2005data,ayanzadeh2019quantum,ayanzadeh2019sat,mousavi2019survey}; therefore, one needs to apply convex (or non-convex) relaxations or greedy algorithms for efficient sparse recovery in the realm of classical computing \cite{foucart2013mathematical,rani2018systematic,mousavi2019survey}.
It is possible to cast the original problem of compressive sensing, shown in \eqref{eqn:cs_ell_0}, to the Boolean satisfiability problem (SAT) \cite{ayanzadeh2019sat} and take advantage of modern SAT solvers; however, SAT-based compressive sensing requires notably more computational resources, compared to the convex optimization methods (e.g., the $\ell_1$-norm sparse recovery techniques).

Restricting the elements of $\mathbf{x}$ to take their values from $\{0,1\}$ leads to a discrete optimization problem—so-called binary compressive sensing (BCS)—that is more challenging, compared to the standard (i.e., continues) compressive sensing \cite{ayanzadeh2019quantum,ayanzadeh2019sat}.
As an illustration, we cannot directly apply currently available greedy algorithms to recover sparse binary (and more generally discrete) sparse signals. In the same way, we need to embed additional constraints to adopt the convex optimization techniques (e.g., the $\ell_1$-norm minimization) for the recovery of sparse binary/discrete signals.

The standard compressive sensing assumes that measurements come from noiseless sources; nevertheless, such an assumption is invalid in real-world applications.
For handling noisy measurements, we can apply the idea of penalty methods and reformulate problem \eqref{eqn:cs_ell_0}, for binary signals, as
\begin{equation}
	\label{eqn:BCS_LASSO}
	\min_{\mathbf{x} \in \{0,1\}^N} {\|\mathbf{y} – A\mathbf{x}\|_{2}^{2} + \lambda \|\mathbf{x}\|_0},
\end{equation}
where the penalty parameter $\lambda \in [0,+\infty)$ controls the trade-off between the feasibility and sparsity of solutions \cite{mousavi2019survey,ayanzadeh2019quantum}.

We can employ quantum annealers to directly address the $\ell_0$-norm problem of BCS, shown in \eqref{eqn:BCS_LASSO} \cite{ayanzadeh2019quantum,ayanzadeh2020leveraging}. 
In practice, nevertheless, technological barriers in manufacturing physical quantum annealers (like noise and decoherence) reduce the recovery accuracy \cite{ayanzadeh2020reinforcement}.
Furthermore, the performance of the quantum annealing based BCS is significantly sensitive to the optimality of the penalty parameter that balances the feasibility and sparsity of results, and finding an optimum penalty parameter is nontrivial \cite{ayanzadeh2020leveraging}.
In this study, we leverage the idea of a statistical ensemble to advance quantum annealing based binary compressive sensing. 

\section{Quantum Annealing based BCS} \label{sec:QBCS}
Quantum annealing (QA)  is a meta-heuristic that applies adjustable quantum fluctuations into a problem and can outperform thermal annealing, a.k.a. simulated or classical annealing \cite{kadowaki1998quantum,das2008colloquium,ayanzadeh2020reinforcement}.
Quantum annealers are a type of adiabatic quantum computers that can sample from the ground state(s) of a given Ising Hamiltonian at cryogenic temperatures in near-constant time \cite{kadowaki1998quantum,ayanzadeh2019quantum_assisted,ayanzadeh2020reinforcement}. 
For instance, the D-Wave quantum annealers receive coefficients of a quadratic unconstraint binary optimization (QUBO) form, as an executable quantum machine instruction (QMI), and returns the ground state of the following quadratic objective function:
\begin{equation}	
	\label{eqn:qubo_energy}
	E_{\mathrm{QUBO}}{(\mathbf{x})} = \sum_{i \leq j}^{N}{\mathbf{x}_iQ_{ij}\mathbf{x}_j},
\end{equation}
where  $\mathbf{x} \in \{0,1\}^N,$ $N$ denotes the number of quantum bits (qubits), and diagonal and off-diagonal entries of ${Q}$ represent linear and quadratic coefficients, respectively \cite{ayanzadeh2020leveraging}.

To solve a problem on a D-Wave quantum processor, therefore, one needs to define a QUBO form (or its equivalent Ising Hamiltonian) whose ground state represents the optimum solution for the original problem of interest \cite{ayanzadeh2020leveraging,ayanzadeh2020sat++}.
In our previous work \cite{ayanzadeh2019quantum}, we showed how to cast problem \eqref{eqn:BCS_LASSO} to \eqref{eqn:qubo_energy} via:
\begin{equation}
	\label{eqn:QBCS_h}
	{Q}_{ii} = \lambda + \sum_l{ A_{li} \left({-2\mathbf{y}_l + A_{li} }\right)}
\end{equation}
and
\begin{equation}
	\label{eqn:QBCS_J}
	Q_{ij} = 2\sum_l{ A_{li}A_{lj} }.
\end{equation}

\section{Ensemble QA-based BCS} \label{sec:method}
Although quantum annealers can draw samples from the ground state(s) of a given Ising Hamiltonian in near-constant time, the current generation of the quantum annealers have limitations that not only restrict the process of mapping problems into an executable quantum machine instruction but also lower the quality of results—including, but not limited to, sparse connectivity of the qubits, noise, decoherence and coefficients’ range/precision limitations \cite{ayanzadeh2020reinforcement,ayanzadeh2020leveraging}. 

In addition, one needs to find a proper value of the penalty parameter $\lambda$ prior to applying Eq. \eqref{eqn:QBCS_h} and \eqref{eqn:QBCS_J} for casting the given BCS problem to a corresponding quantum machine instruction, executable by the quantum annealers. 
In practice, calibrating this penalty parameter is challenging, and can become even intractable \cite{zou2007degrees,candes2007dantzig,bickel2009simultaneous}. 
The penalty parameter $\lambda$ specifies the amount of shrinkage in Eq. \eqref{eqn:BCS_LASSO}. 
When $\lambda \to 0$ the number of eliminated parameters  approaches zero (here, $\|\mathbf{x}\|_0 \to N$). On the other side, when $\lambda \to +\infty$, more parameters are eliminated (here, $\|\mathbf{x}\|_0 \to 0$).

In this study, instead of emphasizing on finding the optimum value for $\lambda$, which can be impractical in many real-world applications, we relax the mapping process to take multiple penalty parameters that are not necessarily optimum. 
Let 
\[
	\Lambda = \{ \lambda^1, \lambda^2, \dots,\}
\]
be the set of different penalty parameters that we use for a given problem, and let 
\[
	H = \{H^1, H^2, \dots \}
\]
denotes the corresponding Ising Hamiltonians that we obtain from applying Eq. \eqref{eqn:QBCS_h} and \eqref{eqn:QBCS_J}.

After executing all (different) corresponding quantum machine instructions for a given problem, we aggregate the resulting samples and look at each element of samples as a binary random variable that follows the Bernoulli distribution. 
Let 
\[
	X=\{\mathbf{x}^1, \mathbf{x}^2, \dots \}
\]
represents the ground states of corresponding Hamiltonians in $H$, attained by a quantum annealer.
We can adopt the idea of ensemble quantum annealing \cite{ayanzadeh2020ensemble} and estimate the optimum solution, denoted by $\tilde{\mathbf{x}},$ as
\begin{equation}
	\label{eqn:EQA_x}
	\tilde{\mathbf{x}}_i=\left[{\frac{1}{n}\sum_{j=1}^n{\mathbf{x}_i^j}} \right], \quad\quad \textnormal{for}\; i=1,2,\dots, N,
\end{equation}
where $n$ denotes the number of recoveries in $X.$
Since we assume that the sparsest solution of the given BCS problem is unique, when $|H| \to +\infty$, we can expect that the majority of the ground states be identical to the sparsest solution of the original problem of interest.

\section{Experiment Results} \label{sec:results}
We employed a D-Wave 2000Q quantum processor (located at Burnaby, British Columbia) for running our experiments.
The current generation of the D-Wave quantum annealers includes more than 2,000 qubits; nevertheless, owing to the sparse connectivity of qubits, they are limited to cliques of size at most 63. 
Hence, in this study, we used random benchmark BCS problems of size $N=60$ \cite{ayanzadeh2019quantum,ayanzadeh2020leveraging}.
The problem set includes 50 random 5-sparse binary signals with corresponding measurement vectors and coding matrices for $m=30, 40$ and $50.$
To avoid the impact of embedding (i.e., chaining multiple physical qubits for representing virtual qubits with higher connectivity) in our evaluations, for all test instances, we used a fixed embedding of a clique of size 60 on the current working graph of the D-Wave QPU. In the same manner, we set the chaining-strength of all problem embeddings to 1.5.

In this experiment, we requested for 1,000 samples/reads for all QMIs. 
After retrieving raw samples from a D-Wave QPU, we performed the majority voting scheme for remediating broken chains. 
We also applied SQC \cite{dorband2018method,ayanzadeh2020leveraging,ayanzadeh2020post_quantum}, as a post-quantum error correction scheme, on all samples and used the best sample (sample with lowest energy value) as the recovered binary signal, denoted by $\tilde{\mathbf{x}}$.
For every recovery, we used 
\[
	e=\frac{\left\| \mathbf{x} -\tilde{\mathbf{x}} \right\|_2^2}{N}
\]
to measure the recovery error. 
Table \ref{tbl:QABCS_error} displays minimum, maximum, average and variance of recovery errors for QA-based BCS with different penalty parameters ($\lambda$), and compares it with the proposed ensemble QA-based BCS, where $\Lambda=\{12,14,20\},$ for $m=30, 40$ and $50.$
Table \ref{tbl:QABCS_sparsity} presents the sparsity rate of the recovered 5-sparse binary signals.
Experiment results reveal that the proposed ensemble QA-based compressive sensing is notably less sensitive to the calibration of the penalty parameter.

\begin{table}[!t]
	\center
	\caption{
Minimum, maximum, average and variance of recovery errors using QA-based BCS (with $\lambda=12, 14, 16, 18, 20$) and ensemble QA-based BCS (with $\Lambda=\{12, 16, 20\}$). 
	}
	\label{tbl:QABCS_error}
	\begin{tabular}{clcccc} 
	$m$	&	penalty		&	min	&	max	&	mean	&	var($10^{-4}$)\\ 	
	\hline
	\multirow{6}{*}{30}
	&	$\lambda=12$		&	0	&	0.12	&	0.044	&	7.0\\
	&	$\lambda=14$		&	0	&	0.10	&	0.039	&	6.3\\
	&	$\lambda=16$		&	0	&	0.08	&	0.036	&	6.2\\
	&	$\lambda=18$		&	0	&	0.10	&	0.037	&	6.2\\
	&	$\lambda=20$		&	0	&	0.08	&	0.039	&	5.6\\
	&	$\Lambda=\{12,16,20\}$		&	0	&	0.08	&	0.036	&	6.2\\
	\hline
	\multirow{6}{*}{40}
	&	$\lambda=12$		&	0	&	0.07	&	0.016	&	2.4\\
	&	$\lambda=14$		&	0	&	0.07	&	0.015	&	2.9\\
	&	$\lambda=16$		&	0	&	0.07	&	0.015	&	2.8\\
	&	$\lambda=18$		&	0	&	0.07	&	0.016	&	3.2\\
	&	$\lambda=20$		&	0	&	0.10	&	0.021	&	4.3\\
	&	$\Lambda=\{12,16,20\}$		&	0	&	0.07	&	0.014	&	2.9\\
	\hline
	\multirow{6}{*}{50}
	&	$\lambda=12$		&	0	&	0.05	&	0.006	&	1.3\\
	&	$\lambda=14$		&	0	&	0.03	&	0.005	&	0.9\\
	&	$\lambda=16$		&	0	&	0.03	&	0.005	&	0.9\\
	&	$\lambda=18$		&	0	&	0.03	&	0.005	&	1.0\\
	&	$\lambda=20$		&	0	&	0.03	&	0.006	&	1.1\\
	&	$\Lambda=\{12,16,20\}$		&	0	&	0.03	&	0.005	&	0.9\\
	\end{tabular}
\end{table}

\begin{table}[!t]
	\center
	\caption{
Minimum, maximum, average and variance of sparsity rates using QA-based BCS (with $\lambda=12, 14, 16, 18, 20$) and ensemble QA-based BCS (with $\Lambda=\{12, 16, 20\}$). 
	}
	\label{tbl:QABCS_sparsity}
	\begin{tabular}{clcccc} 
	$m$	&	penalty		&	min	&	max	&	mean	&	var\\ 	
	\hline
	\multirow{6}{*}{30}
	&	$\lambda=12$		&	1	&	11	&	5.42	&	4.80\\
	&	$\lambda=14$		&	1	&	10	&	5.04	&	4.24\\
	&	$\lambda=16$		&	1	&	9	&	4.70	&	3.93\\
	&	$\lambda=18$		&	1	&	8	&	4.28	&	3.92\\
	&	$\lambda=20$		&	0	&	8	&	3.90	&	3.65\\
	&	$\Lambda=\{12,16,20\}$		&	1	&	9	&	4.70	&	3.93\\
	\hline
	\multirow{6}{*}{40}
	&	$\lambda=12$		&	3	&	7	&	5.00	&	1.00\\
	&	$\lambda=14$		&	3	&	7	&	4.82	&	0.87\\
	&	$\lambda=16$		&	3	&	6	&	4.72	&	0.84\\
	&	$\lambda=18$		&	2	&	6	&	4.50	&	0.89\\
	&	$\lambda=20$		&	1	&	6	&	4.08	&	1.19\\
	&	$\Lambda=\{12,16,20\}$		&	3	&	6	&	4.70	&	0.81\\
	\hline
	\multirow{6}{*}{50}
	&	$\lambda=12$		&	4	&	6	&	5.18	&	0.19\\
	&	$\lambda=14$		&	4	&	6	&	5.08	&	0.15\\
	&	$\lambda=16$		&	4	&	6	&	5.06	&	0.18\\
	&	$\lambda=18$		&	4	&	6	&	5.00	&	0.16\\
	&	$\lambda=20$		&	4	&	6	&	4.98	&	0.18\\
	&	$\Lambda=\{12,16,20\}$		&	4	&	6	&	5.06	&	0.18\\
	\end{tabular}
\end{table}

\section{Discussion} \label{sec:discussion}
The original problem of compressive sensing in sparse recovery (i.e., the $\ell_0$-norm sparse recovery) is NP-hard and restricting the elements of the original (sparse) signal to take their values from $\{0,1\}$ leads to a discrete optimization problem—so-called binary compressive sensing (BCS)—that is significantly more challenging, compared to the standard (continues) compressive sensing. 
One can cast the original problem of BCS to minimize a quadratic unconstrained binary optimization (QUBO) form, which is tractable by quantum annealers, whose ground state represents a solution to the given problem of BCS. 

The performance of the sparse recovery in QA-based BCS is highly sensitive to the penalty parameter that balances the feasibility and sparsity of recoveries.
Calibrating the penalty parameter is nontrivial—in several cases, it can become intractable.
In this study, hence, we introduced the idea of ensemble QA-based BCS that leverages the idea of the statistical ensemble to improve the quality of QA-based BCS. 
Since executing quantum machine instructions on the quantum annealers can result in an excited state, rather than the ground state of the given Ising Hamiltonian, we use different penalty parameters to generate multiple distinct QUBOs whose ground state(s) represent a potential solution of the original problem. 
We then employ the attained samples from minimizing all corresponding (different) QUBOs to estimate the solution of the original problem of binary compressive sensing. 
Our experiments, on a D-Wave 2000Q quantum processor, demonstrated that the proposed ensemble quantum annealing approach is significantly less sensitive to the calibration of the penalty parameter $\lambda$. 
It is worth highlighting that the uniqueness of the (sparse) solution is necessary for a successful recovery in our proposed ensemble QA-based BCS.

\bibliographystyle{IEEEbib}
\bibliography{biblio}

\end{document}